\documentclass[fleqn,usenatbib]{mnras}

\usepackage{newtxtext,newtxmath}

\usepackage[T1]{fontenc}

\DeclareRobustCommand{\VAN}[3]{#2}
\let\VANthebibliography\thebibliography
\def\thebibliography{\DeclareRobustCommand{\VAN}[3]{##3}\VANthebibliography}

\usepackage{graphicx,graphics}	
\usepackage{amsmath}	
\usepackage{enumitem}
\usepackage{caption}
\usepackage{booktabs}
\usepackage{varwidth}
\usepackage{tcolorbox}
\usepackage{caption}
\usepackage{natbib}
\usepackage{wasysym}

\graphicspath{{figs/}}

\newenvironment{tightcenter}{%
	\setlength\topsep{0pt}
	\setlength\parskip{0pt}
	\begin{center}
}{%
 	\end{center}
}

\newcommand{\angstrom}{\textup{\AA}}

\newcommand\chandra{{\it Chandra}}

\newcommand\sherpa{{\it Sherpa}}

\newcommand\alphaox{\alpha_{\rm ox}}

\newcommand\gal{{J1429}}
\newcommand{\nh}{$N_{\rm H}$}

\newcommand{\xmm}{{XMM-{\it Newton}}}

\newcommand{\ms}{$\rm M_{\odot}$}

\newcommand{\fluxcgs}{erg~s$^{-1}$~cm$^{-2}$}
\newcommand{\lumcgs}{erg~s$^{-1}$}

\title[X-ray luminous RL quasar at $z=6.18$]{The extremely X-ray luminous radio-loud quasar CFHQS J142952+544717 at $z=6.18$ under Chandra high-angular resolution lens}

\author[G.~Migliori et al.]
{G.~Migliori$^{1}$\thanks{E-mail: giulia.migliori@inaf.it}, 
A.~Siemiginowska$^{2}$, M.~Sobolewska$^{2}$, C.~C.~Cheung$^{3}$, {\L}.~Stawarz$^{4}$, D.~Schwartz$^{2}$, B.~Snios$^{2}$,
\newauthor 
 A.~Saxena$^{5,6}$, V.~Kashyap$^{2}$
\\
$^{1}$Istituto di Radioastronomia - INAF, Via P. Gobetti 101, I-40129 Bologna, Italy\\
$^{2}$Center for Astrophysics $|$ Harvard \& Smithsonian, 60 Garden St., Cambridge, MA, 02138, USA\\
$^{3}$Space Science Division, Naval Research Laboratory, Washington, DC 20375, USA\\
$^4$Astronomical Observatory, Jagiellonian University, ul. Orla 171, 30-244 Krak\'ow, Poland\\
$^5$Department of Physics, University of Oxford, Denys Wilkinson Building, Keble Road, Oxford OX1 3RH, UK\\
$^6$Department of Physics and Astronomy, University College London, Gower Street, London WC1E 6BT, United Kingdom
}

\date{Accepted XXX. Received YYY; in original form ZZZ}

\pubyear{2023}

\begin{document}
\label{firstpage}
\pagerange{\pageref{firstpage}--\pageref{lastpage}}
\maketitle

\begin{abstract}
We present the first X-ray observation at sub-arcsecond resolution of the high-redshift ($z=6.18$) radio-loud quasar CFHQS J142952+544717 (\gal). The $\sim 100$ net-count 0.3--7\,keV spectrum obtained from $\sim 30$\,ksec \chandra{} exposure is best fit by a single power-law model with a photon index $\Gamma=2.0\pm0.2$ and no indication of an intrinsic absorber, implying a 3.6--72\,keV rest-frame luminosity $L_{\rm X}=(2.3^{+0.6}_{-0.5})\times10^{46}$\,\lumcgs. We identify a second X-ray source at 30\arcsec\, distance from \gal{} position, with a soft ($\Gamma\simeq 2.8$) and absorbed (equivalent hydrogen column density $N_{\rm H} <13.4\times 10^{20}$\,cm$^{-2}$) spectrum, which likely contaminated J1429 spectra obtained in lower angular resolution observations. 
Based on the analysis of the \chandra{} image, the bulk of the X-ray luminosity is produced within the central $\sim 3$\,kpc region,
either by the disk/corona system, or by a moderately aligned jet. In this context, we discuss the source properties in comparison with samples of low- and high-redshift quasars. We find indication of a possible excess of counts over the expectations for a point-like source in a 0.5\arcsec--1.5\arcsec\, ($\sim 3-8$\,kpc) annular region. The corresponding X-ray luminosity at J1429 redshift is $4\times 10^{45}$\,\lumcgs. If confirmed, this emission could be related to either a large-scale X-ray jet, or a separate X-ray source. 

\end{abstract}




\section{Introduction}
The formation and growth of early black holes and their impact on the evolution of structures across the Universe is at the forefront of current astrophysical research. One of the main open problems relates to the formation and evolution of radio sources and the significance of radio phenomena (i.e., jets and lobes) produced by growing black holes in the feedback and co-evolution of galaxies and clusters of galaxies. We still do not understand why only a small fraction of quasars exhibits powerful radio emitting structures extending to large, in some cases even Mpc, scales. And yet, the existence of jetted quasars at high redshift  challenges models of structure formation, as their radio power requires a very massive black hole, $\rm M_{bh} >10^9- 10^{10} \rm M_{\odot}$ \citep[e.g.,][and references therein]{Croton2006,Volonteri2009,Valiante2016}.

The energy released by the jet into the interstellar medium (ISM) may impact the evolution of the host galaxy \citep{Fragile2004,Gaibler2012,Mukherjee2018,Meenakshi2022}. Observational evidence of such effect is still limited and its interpretation controversial, with positive and negative feedback considered to play a role. \citep[e.g.][]{Bicknell2000,Croft2006,Nesvadba2010,Salome2015,Lanz2016,Nesvadba2020,Girdhar2022}.  
In a recent work, \citet{PCC2023} found indication that the supermassive black holes (SMBHs) powering radio-loud\footnote{Here we assume the classical separation between radio-loud (RL) and radio-quiet (RQ) quasar based on the rest-frame radio to optical flux density ratio $R$, with the radio being measured at 5\,GHz and the optical at 4400 \AA\, \citep{Kellerman1989}. The divide is set at $R=10$.} active galactic nuclei (AGN) in the $0.3<z<4$ range, are overall more massive than what expected by the scaling relation between the masses of SMBHs and their host spheroids in the local Universe. The proposed explanation involves a relevant role of ``radio-mode'' AGN feedback, leading to a rapid growth of SMBHs at early epochs while influencing the star-formation history of the AGN host galaxy \citep[see also][]{Jolley2008,Diana2022}. 

In order to investigate radio-mode feedback in the AGN evolution, sizeable samples of radio-loud AGN at high redshift are needed. Indeed, the number of known high-redshift radio sources ($z>5$)  has increased significantly during the past several years \citep[e.g][for a general review]{Banados2015,Banados2018,Fan2022}, with a rapid sequence of record breaking discoveries \citep[][to name a few]{Willott2010,Saxena2018,Belladitta2020,Banados2021,Connor2021,Endsley2022,Ighina2023}.
At the time of writing this article, there are 10 radio quasars known at $z>6$. In addition, \citet{Gloudemans2022}  report the discovery of 24 radio-bright (21 radio-loud) quasars at $4.9 \lesssim z \lesssim 6.6$ by combining DESI Legacy Imaging Surveys and LOFAR Two-metre Sky Survey (LoTSS), while two more targets selected from the Rapid ASKAP Continuum Survey (RACS) and the Dark Energy Survey (DES) have been spectroscopically confirmed at $z\sim6.1$ \citep{Ighina2022c}.

X-ray observations provide important constraints on the physical processes associated with the accretion onto a SMBH. X-ray selected samples are key to investigate the accretion history of AGN free from absorption biases that affect other wavelengths \citep[][and references therein]{Wolf2021,Barlow2023}. X-ray observations can also help to constrain radiative processes at work in quasar jets and help to infer their physical properties. 
For example, the contribution of the inverse-Compton scattering off the Cosmic Microwave Background photons by relativistic electrons \citep[IC/CMB; see][]{Tavecchio2000,Celotti2001} to the X-ray luminosity of jets in the local Universe, is highly debated \citep[e.g.,][]{Stawarz04,Hardcastle2006,Meyer2014,Breiding2023}. Importantly, the IC/CMB component should be more easily observable in high-$z$ jets because of the $(1+z)^4$ dependence of the CMB photon density \citep{Schwartz2002}. Indeed, the IC/CMB model appears to account well for the emission of several high-$z$ jets detected by the \chandra{} X-ray observatory \citep{Siemiginowska2003,Cheung2006,Cheung2012,Simionescu2016,Wu2017,Napier2020,Snios2022,Ighina2022c} even though, thus far, sample studies have not provided robust indication of the expected X-ray emission enhancement with redshift \citep{Wu2013,McKeough2016,Ighina2019,Ighina2021}, a major limitation to this test being the paucity of radio-loud quasars known at high redshifts ($z>5$) and the even smaller number of X-ray detected radio quasars.

Of the $z>6$ radio quasars known to date \citep[see][for a recent compilations]{Momjian2018,Liu2021,Ighina2023}, currently only three have reported X-ray detections \citep[][]{Khorunzhev2021}. Of these, CFHQS J142952+544717 (hereafter, \gal) is
a remarkable source in light of its high X-ray luminosity \citep[exceeding $10^{46}$\,\lumcgs in the 2-10\,keV rest-frame energy band,][]{medvedev20,medvedev21} and its radio properties suggesting a young radio phase \citep{Frey2011}.
   
Indeed, the source was part of the sample of high-$z$ candidate young radio sources selected by our group for \chandra{} observations with the goal of investigating the high-energy properties and the evolution of newly born radio jets. The first part of the sample was presented in \citet{Snios2020}, while a publication on the remaining targets is in preparation. Here, we present the results of the first X-ray study at arc-second resolution of \gal. In the $\sim 2.2-50$\,keV rest-frame band covered by \chandra{}, different radiative processes could be at work. Broadly summarizing, these can be  related either the AGN or the radio structures (jets and lobes) of \gal. Comptonization of the ultraviolet (UV) disk photons by the electrons in a hot (10$^{8-9}$\,K) corona \citep{Haardt1993} surrounding the SMBH produces a power-law with a characteristic roll-over at energies of a few hundred keV. Non-thermal X-ray emission in the extended radio structures is produced via IC/CMB but also via IC of the jet synchrotron photons and off the nuclear photons, which include direct UV disk photons or disk photons reprocessed in the broad line regions (BLR) or in the torus \citep[see also the discussion in][]{medvedev21}.  

The paper is organized as follows: after summarizing the main information on \gal{} (Sec\,\ref{sec:target}), we present the \chandra{} observation and X-ray analysis results in Sec.\,\ref{sec:obs}. The source properties are discussed in Sec.\,\ref{sec:disc} and we draw our conclusions in Sec.\,\ref{sec:conc}.

\subsection{CFHQS J142952+544717}
\label{sec:target}

\begin{figure}
	\includegraphics[trim=0.8cm 7.cm 0.8cm 7.5cm, clip=true,width=\columnwidth]{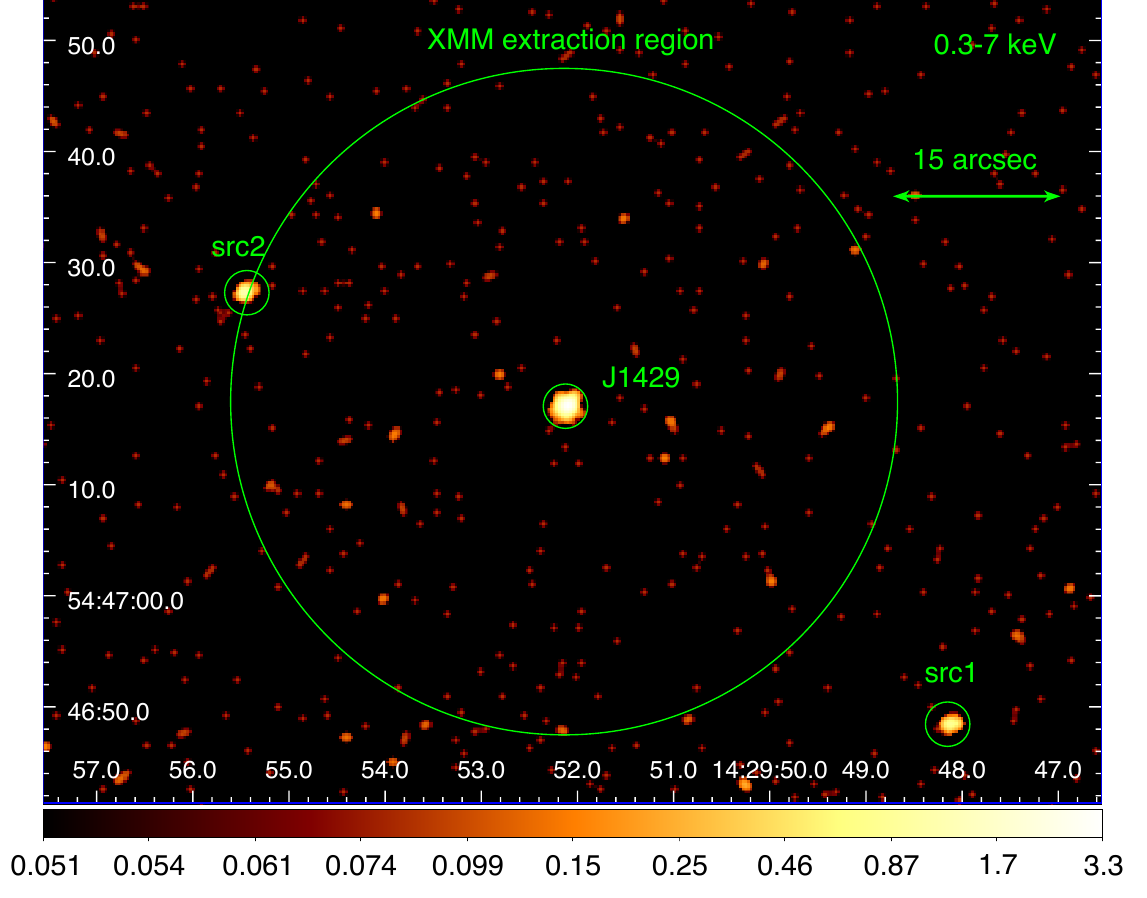}
	\includegraphics[trim=0.8cm 7.cm 0.8cm 7.5cm, clip=true,width=\columnwidth]{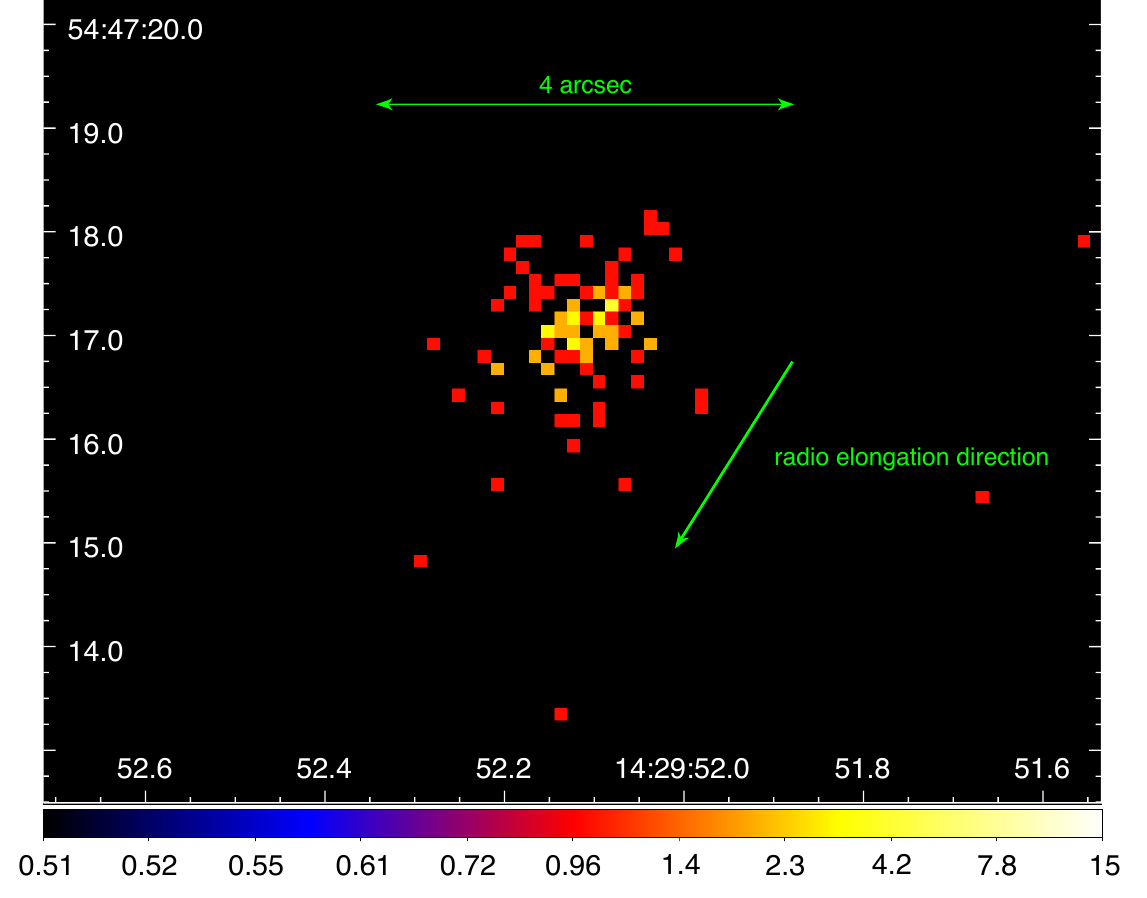}
    \caption{Upper panel: \chandra{} 0.3--7.0\,keV image of \gal{} field. The pixel size is set to half the original ACIS pixel (0.246\arcsec/pix). The image was smoothed using a Gaussian function with $\sigma= 1.5$. The large circle corresponds to the 30\arcsec\, radius extraction region for the PN spectrum in \citet{medvedev21}.
    Lower panel: zoom view on \gal{} displayed at the high resolution binning of 0.123\arcsec/pix. The green diagonal line indicates the direction of the elongation of the radio structure as seen in the 1.6\,GHz EVN map presented in \citet{Frey2011}. The color scales are logarithmic.
    }
    \label{fig:chandra}
\end{figure}

J1429 was first observed spectroscopically as part of the Canada-France High-$z$ Quasar Survey \citep[CFHQS,][]{Willott2005}. The quasar redshift is taken to be $z=6.1837 \pm 0.0015$ (hereafter, $z=6.18$), as determined from the CO (1-0) line emission \citep{Wang2011}. Its absolute magnitude ($M_{1450}=-25.85$) places the source at the bright end of the quasar luminosity function at $z\sim 6$ \citep{Willott2010}. It is classified as a radio-loud quasar based on a reported radio loudness parameter $R=109\pm 9$ \citep[see][]{Banados2015}.
\gal\ was repeatedly observed in the optical band over more than a decade without displaying large amplitude variability. 
Radio observations cover the 120 MHz to 32\,GHz band providing a good characterization of the radio spectrum, which is flat below 5\,GHz and steepens (up to $\alpha\sim 1.0$) at higher frequencies. The steep spectrum, lack of strong variability, and VLBI observations
showing a compact ($< 100$\,pc) but marginally resolved morphology \citep{Frey2011}, make \gal\ a candidate young radio source. 
 \citet{Frey2011} imaged \gal\ with the European VLBI Network (EVN) at 1.6 and 5 GHz and noted a faint extension to the SE in the 1.6 GHz map only (see Table 1 therein).
The extension is at a position angle PA$=138\deg$  (indicated in Figure \ref{fig:chandra}, bottom) and is offset from the putative radio core by 6.4 mas (36 pc, projected).
The intrinsic compactness of the radio structure is also supported by the measured brightness temperature ($T_{\rm B}\simeq 10^9$\,K), which disfavours Doppler-boosted radio emission \citep{Frey2011}.

Observations at 32\,GHz \citep{Wang2011} and 250\,GHz \citep{Omont2013} investigated the host galaxy properties unveiling vigorous star formation and possibly the presence of a companion galaxy at $\sim 7$\,kpc distance. These results found support in recent NOEMA observations of [CII] line emission and of the underlying continuum \citep{Khusanova2022}, which constrained the star-formation rate as SFR\,$=520-870$\,\ms\,yr$^{-1}$. The [CII] line 
properties can be explained with two merging galaxies or, alternatively, with an AGN-driven outflow.

In X-rays, \gal{} was first detected by the extended ROentgen Survey with an Imaging Telescope Array (eROSITA) in December 2019 \citep{medvedev20}. The short (160 s) exposure was sufficient to measure a 0.3--2\,keV flux of $\sim 8 \times 10^{-14}$\,\fluxcgs\ and a rest-frame 2-10\,keV luminosity of a few $10^{46}$\,\lumcgs, that makes \gal\ one of the most X-ray luminous high-$z$ quasars known to date. A follow-up 20\,ksec \xmm\ director's discretionary time observation on 2020 July 24 improved the count statistics of the X-ray spectrum, which is best-fit by a power-law with a steep photon index, $\Gamma=2.5\pm0.2$ and a moderate level of absorption, \nh\,$=(3 \pm 2) \times 10^{22}$\,cm$^{-2}$ at the source redshift \citep{medvedev21}. The source did not display any significant flux variability among the eROSITA and \xmm\ observations, separated by $\sim 7.5$ months.

We adopt a flat $\Lambda$CDM cosmological model with $h= 0.70$ and $\Omega_{\Lambda}=0.7$. The source redshift, $z=6.18$, corresponds to a luminosity distance of 59.8 Gpc and a scale of 5.6\,kpc/\arcsec. The observed 0.3--7\,keV energy band translates into a 2.2--50.3\,keV rest frame band.

\section{X-ray Observations}
\label{sec:obs}
\gal\ was included in the sample of approved high redshift \chandra\ targets (Cycle AO21, PI: Siemginowska) selected from \cite{Coppejans2016,Coppejans2017} catalog of radio-loud AGN with redshifts above $z>4.5$. The selection was based on the shape of radio spectra and available morphology in the VLBI observations. The selected sources have in particular steep or peaked radio spectra, likely dominated by compact radio lobes rather than relativistically beamed jets \citep{Readhead1996a,O'Dea1998}. 
Radio morphologies confirm the presence of double or single extended structures on relatively small scales (a few kpc), suggesting that these sources have not grown to large-scale radio galaxies either because they are young, or because of the impact of the environment preventing their growth \citep[see][for a review]{odea2021}. The analysis of the full sample will be presented in a forthcoming paper.

\begin{table*}
	\caption{\chandra\ best fit spectral models.
	\label{tab:results}}
	\begin{tightcenter}
	\footnotesize
	\begin{tabular}{l c c c c c c c}
		\hline \hline
		 Object & 0.3--7\,keV  &0.3--2\,keV  &2--7\,keV  &\nh  &$\Gamma$ &cstat  &Flux\\
		        &counts &counts &counts &cm$^{-2}$ & & &($\times$10$^{-14}$\,\fluxcgs)\\
		 (1) &(2)  &(3)  &(4)  &(5)  &(6)  &(7) &(8) \\ 
		\hline
		J1429  &96.8$\pm$9.9   &59.3$\pm$7.7    &37.5$\pm$6.1    &\nh$_{\rm ,\, Gal}$ (f)    &2.0$\pm$0.2 &159/148    &$5.4^{+1.4}_{-1.2}$ \\ 
		src1   &20.4$\pm$4.6   &12.8$\pm$3.6    &7.6$\pm$2.8    &\nh$_{\rm ,\, Gal}$ (f)     &2.1$\pm$0.5  &93/92  &$1.2^{+1.0}_{-0.5}$  \\ 
		src2   &25.3$\pm$5.1   &15.4$\pm$4.0    &9.7$\pm$3.2    &$<1.3 \times 10^{22}$   &2.8$\substack{+1.4\\-1.3}$    &98/116 &$4.0^{+2.1}_{-2.0}$  \\
		\hline
	\end{tabular}
    \end{tightcenter}
    {(1) Object name; (2), (3) \& (4) Total, soft and hard net X-ray counts; (5) column density in units of $10^{20}$\,cm$^{-2}$; (f) fixed at the Galactic value \nh$_{\rm ,\, Gal}=1.15 \times 10^{20}$\,cm$^{-2}$; (6) power-law photon index; (7) Cash statistic; (8) 0.5--10\,keV unabsorbed, observer-rest-frame flux.}
\end{table*}

\subsection{Chandra data reduction \& analysis}
\label{sec:analysis}

The 30.56\,ksec \chandra\ observation was performed on 2021-08-03 using the Advanced CCD Image Spectrometer (ACIS-S) with the readout of the full CCD in the VFAINT mode.  The target was located on the S3 chip at Y$=-0.1$\,arcmin off-set from the nominal aim-point.  
We used the CIAO v.4.14 \citep{Fruscione2006} software for the X-ray data analysis and reprocessed the data using {\tt chandra\_repro} script to apply the most recent calibrations (CALDB v.4.9.8) and the sub-pixel adjustment algorithm ({\tt pix\_adj=EDSER}) for the best angular resolution of the image.

A visual inspection of the 0.3--7\,keV image in {\tt ds9} clearly shows the presence of a point-like source at the optical coordinates of \gal\ (see Figure \ref{fig:chandra}). The X-ray centroid is located at the coordinates (J2000)\,RA=14:29:52.12 DEC=+54:47:16.99.  The \chandra\ image confirms the presence of a second X-ray source (src1) at 44$\arcsec$.8 south-west of \gal, already identified in \citet{medvedev21}. In addition, it unveils a second field source (src2), at 30$\arcsec$.5 north-east  of \gal, spatially coincident with the infrared source WISEA J142955.47$+$544727.8 ($w1=17.96\pm0.18$\,mag, $w2>17.90$\,mag, $w3>12.91$\,mag, $w4>9.40$\,mag). 

We used the {\tt ds9 DAX} aperture photometry to extract the net counts in the full (0.3--7\,keV), soft (0.3--2\,keV) and hard (2.0--7.0\,keV) energy bands of \gal\ and of the two field sources from circular regions with $\sim 2\arcsec$ radius
(95\% PSF ECF at 1.5\,keV) centered on the respective X-ray centroids. 
The same regions were used to extract the spectra and response files of the three sources with {\tt specextract}. The background spectrum was extracted from a region surrounding the source and free of known point sources. 
All model fitting was performed in \sherpa\, \citep{Freeman2001} assuming the C-statistics based on the Poisson likelihood, and using Nelder-Mead optimization algorithm. Uncertainties are reported at $1 \sigma$ confidence level.

\subsection{Results of Spectral Analysis}
\label{sec:results}

\begin{figure}
	\includegraphics[trim=1.cm 1.cm 1.cm 1.cm, clip=true,width=\columnwidth]{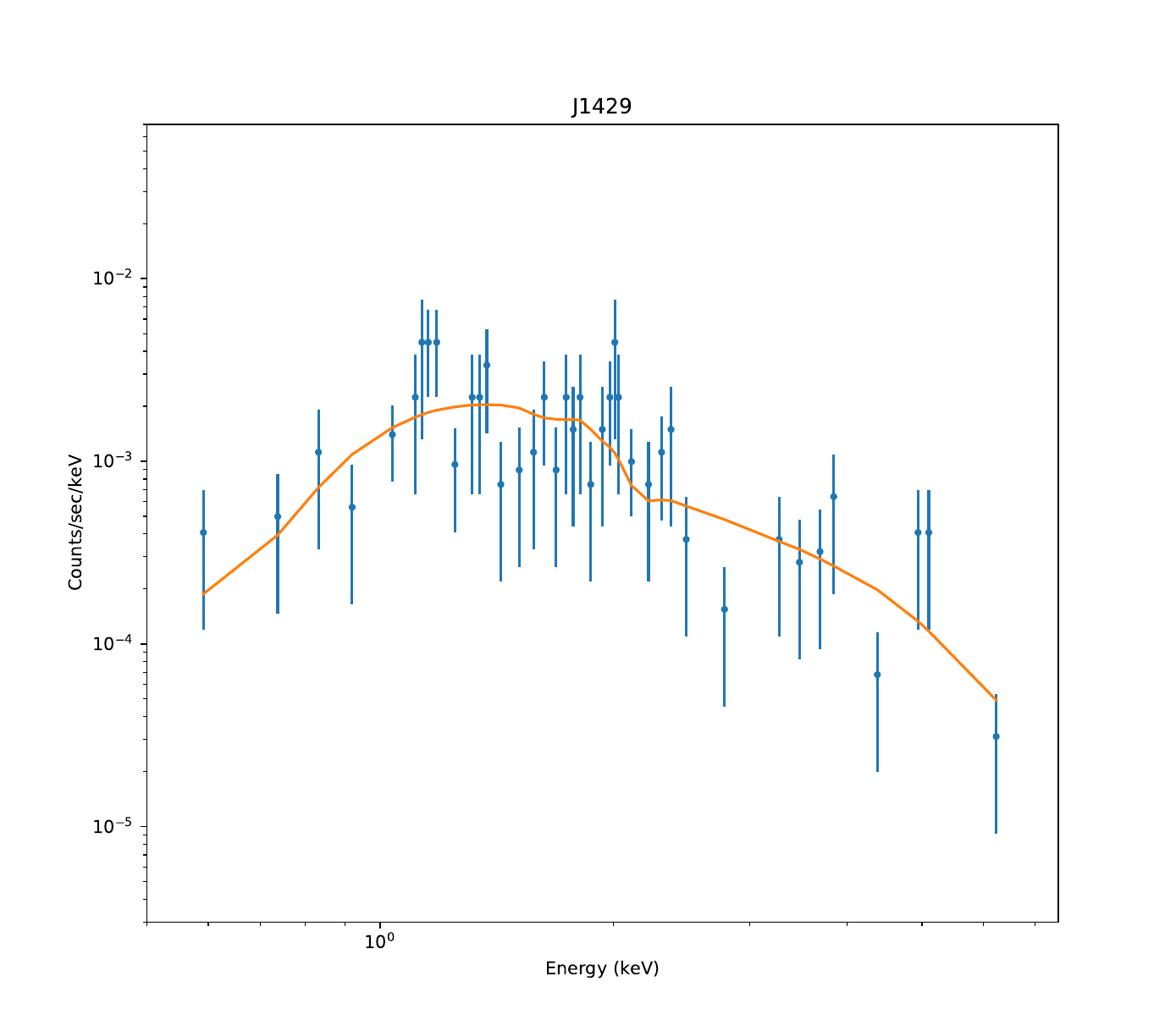}
    \caption{\chandra{} 0.3-7.0 keV spectrum and best-fit absorbed power law model. The spectrum has been rebinned only for visualization purposes.
    }
    \label{fig:chandra_spectrum}
\end{figure}

We assumed an absorbed power-law model and performed a fit to the X-ray spectrum of each source, initially leaving the absorption column free to vary. 
For \gal\ and src1, we did not find any evidence for an absorption parameter value in excess over the measured Galactic column density \citep[\nh$_{\rm ,\, Gal}=1.15\times 10^{20}$\,cm$^{-2}$;][]{nh2016}, while the src2 spectrum appears best modeled assuming a moderately absorbed, steep ($\Gamma\simeq 2.8$) power-law, although the limited statistic provides us only with an upper limit on the intrinsic column density \nh\,$<1.3\times 10^{22}$\,cm$^{-2}$. 
More complex spectral models, including a cut-off power-law, or the addition of a thermal component, do not significantly improve the fit of \gal.
The resulting best-fit model parameters are listed in Table~\ref{tab:results}  and the \gal{} X-ray spectrum and model are shown in Figure \ref{fig:chandra_spectrum}.
For \gal, the best fit photon index is $\Gamma=2.0\pm0.2$ and the 0.5--10\,keV unabsorbed flux is $5.4^{+1.4}_{-1.2}\times 10^{-14}$\,\fluxcgs. 

The results summarized above differ from those of \citet{medvedev21} based on the \xmm\ observation. Their best-fit model includes a steep power-law ($\Gamma= 2.5 \pm 0.2$) with an intrinsic absorber (\nh\,$_{\rm ,\, int}=(3\pm2)\times 10^{22}$\,cm$^{-2}$), although with marginal significance, and gives an unabsorbed 0.2--10\,keV flux of $1.3 \times 10 ^{-13}$\,\fluxcgs{} (the 0.2--10\,keV unabsorbed \chandra{}\ flux of \gal{} being $6.6 \times 10^{-14}$\,\fluxcgs). However, the angular resolution of \xmm\ allowed the authors to identify only one of the two sources in the field of \gal, src1, while src2 remains blended with \gal. 
 Given the \xmm-PN Point Spread Function\footnote{\url{https://xmm-tools.cosmos.esa.int/external/xmm_user_support/documentation/uhb/onaxisxraypsf.html}.}, the two field sources could have contaminated the \xmm{} spectrum of \gal. The spectral analysis indicates that src1 has likely a minimal impact given its greater distance and fainter flux. 
Here, we re-analyze the \xmm\ data in order to evaluate the contribution of src2 emission to the \gal\ spectrum. We followed the standard data reduction and extracted the spectrum from a circular region of $r=30\arcsec$ centered on the position reported in \citet{medvedev21}.

Fitting the \xmm{} spectrum with an absorbed power-law returned values $\Gamma=2.5 \pm 0.2$ and \nh$_{\rm ,\, int}\,=(2^{+2}_{-1})\times 10^{22}$\,cm$^{-2}$, which are fully consistent with the analysis by \citet{medvedev21}. We then tested a composite model, consisting of the sum of two absorbed power laws, and fixed the parameters of one of the two to the best-fit values obtained for \gal. The underlying assumption is that \gal\ has not varied and that the excess flux is due to contamination by src2. In this way, we obtained a moderately absorbed (\nh\,$=1.2^{+1.2}_{-0.7}\times 10^{21}$\,cm$^{-2}$) and very steep ($\Gamma=3.7_{-0.7}^{+1.0}$) power-law with a 0.2--10\,keV absorbed flux of $2.1 \times 10^{-14}$\,\fluxcgs, which is consistent, within uncertainties, with the spectral parameters of src2 in the \chandra{} spectrum. 

We conclude that the contamination from src2 is likely responsible for the softer photon index and higher flux measured in the \xmm\ data. We note that the resulting observed broadband X-ray luminosity $L_{0.1-100\, \rm keV} = 4.2 \times 10^{46}$\,\lumcgs, based on our \chandra\ analysis, is still high, thus \gal{} remains among the most luminous $z>6$ quasars.

\subsection{Image Analysis}

\chandra\ observations result in the highest angular resolution X-ray images to date and allow us to study the X-ray morphology on sub-arcsecond scales.
We performed an analysis of the \gal\ \chandra\ image to understand if the X-ray data are consistent with a point source emission, and also to look for the presence of a diffuse extended component. 
The \chandra\ image of \gal\ contains a relatively small number of counts, $96.8\pm9.9$, however, the background contamination is low (estimated $<1$ count  in the source circular region with $r=2\arcsec$ based on the background surface brightness of 0.054\,cts\,arcsec$^{-2}$). Figure~\ref{fig:chandra} shows the X-ray counts in the quasar region with the 0.123\arcsec pixel size.
No evidence of an extended emission is present beyond 2\arcsec ($\sim 11$\,kpc) radius from the centroid of the X-ray source and we measured a 0.5--7.0\,keV upper limit (90\% confidence limit) of $2.2\times 10^{-15}$\,\fluxcgs. 

Next, we investigated whether the central emission is consistent with a point-like source.
We used CHART\footnote{\url{https://cxc.harvard.edu/ciao/PSFs/chart2/}} \citep{Carter2003} to simulate the \chandra\ PSF centered on the location of the quasar (see the centroid given above). We assumed the input spectrum to be the best fit spectral model with $\Gamma=2.0$ and Galactic absorption, and selected the dither option to match the observation's aspect solution ({\tt asol}) file.  We simulated 500 realizations of the PSF with matched observed exposure time using CHART and projected each of the simulated rays onto the ACIS-S detector using MARX (v.5.2) with the pixel adjustment algorithm setting as pix-adj=EDSER, the same {\tt asol} file for the dither, and the addition of AspectBlur\,=\,0.25\arcsec (see POG and CIAO threads)\footnote{Note that this adds additional 
broadening of the PSF resulting in a broader spread of the photons on the detector; this is a conservative setting for this parameter.}.

\begin{figure}
    \includegraphics[width=\columnwidth]{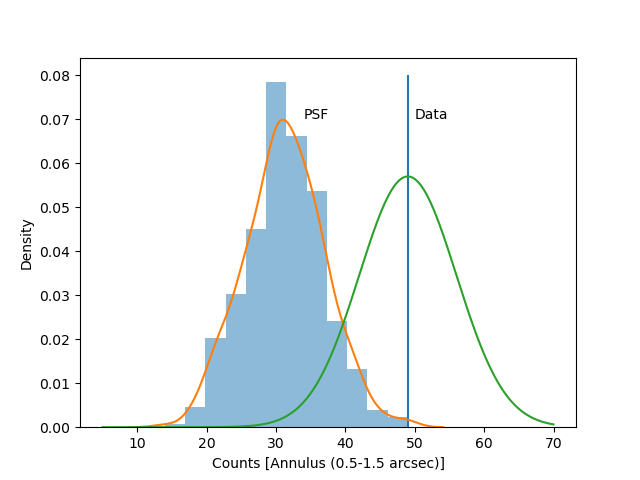}
    \caption{Simulated counts from a point source and the data. The blue histogram shows the distribution of counts in the annulus 0.5--1.5\arcsec overplotted with the KDE curve (orange) for 500 simulations of a point source with 100 counts. The vertical line marks 49 counts detected in the same annulus by \chandra. The green line shows the Gaussian distribution with the mean of 49 and $\sigma=7$ representing the measurement error. The counts in the PSF artefact region were excluded.}
    \label{fig:psfhist}
\end{figure}

For each simulation we calculated the point source counts in an annulus with 0.5\arcsec and 1.5\arcsec radii with the PSF encircled energy fraction at 1\,keV of $\sim 75\%$ and $\sim 95\%$, respectively (calculated from the \chandra\ image in ds9 using DAX). In terms of physical scales we probe the region from 2.8\,kpc to 8.4\,kpc. The distribution of the point source counts is shown in Figure~\ref{fig:psfhist} together with the observed number of counts.
We performed a two sample Kolmogorov-Smirnov (KS) test (using {\tt scipy.ks\_2sampl}) obtaining the p-value\,$<<0.001$, which indicates that the two distributions are different at high confidence level. 
Our analysis points therefore to an excess of X-ray counts with respect to expectations in case of a point-like emission.
We investigated the distribution of the counts, by dividing the annulus in four quadrants. The two north and south-east quadrants have the highest number of counts, however these are basically comparable among each other (11 to 14). As a further test, we extracted the surface brightness profile of the emission in the direction parallel and perpendicular to the observed radio elongation, and compared them with those of the simulated PSF. The observed and simulated orthogonal profiles match each other. The longitudinal profile of \gal\ appears instead to display an asymmetry in the wings, the south-east one being broader, however the difference between the observed and simulated profiles could not be confirmed with the K-S test. 

To summarize, the imaging analysis points to a count excess with respect to a point-source predictions, but we cannot make conclusions on the counts being clustered in a specific location.
Moreover, while compelling, we stress that the result of a count excess should be taken with caution given the relatively low count statistics and the systematic uncertainties in the PSF which cannot be properly included in the simulations \cite[see][for a recent discussion of the \chandra\,PSF uncertainties]{Ma2023}. Indicatively, the excess ($20\pm 13$ counts in excess) corresponds to an unabsorbed flux of $(9\pm6)\times 10^{-15}$\,\fluxcgs\, in the 0.5--7\,keV observed energy band if we assume a power-law model with an intermediate (between radio-loud and radio-quiet AGN) photon index value $\Gamma=1.7$.

\begin{figure*}
	\includegraphics[trim=0.cm 2.cm 0.cm 2.2cm, clip=true,angle=90,width=0.95\columnwidth]{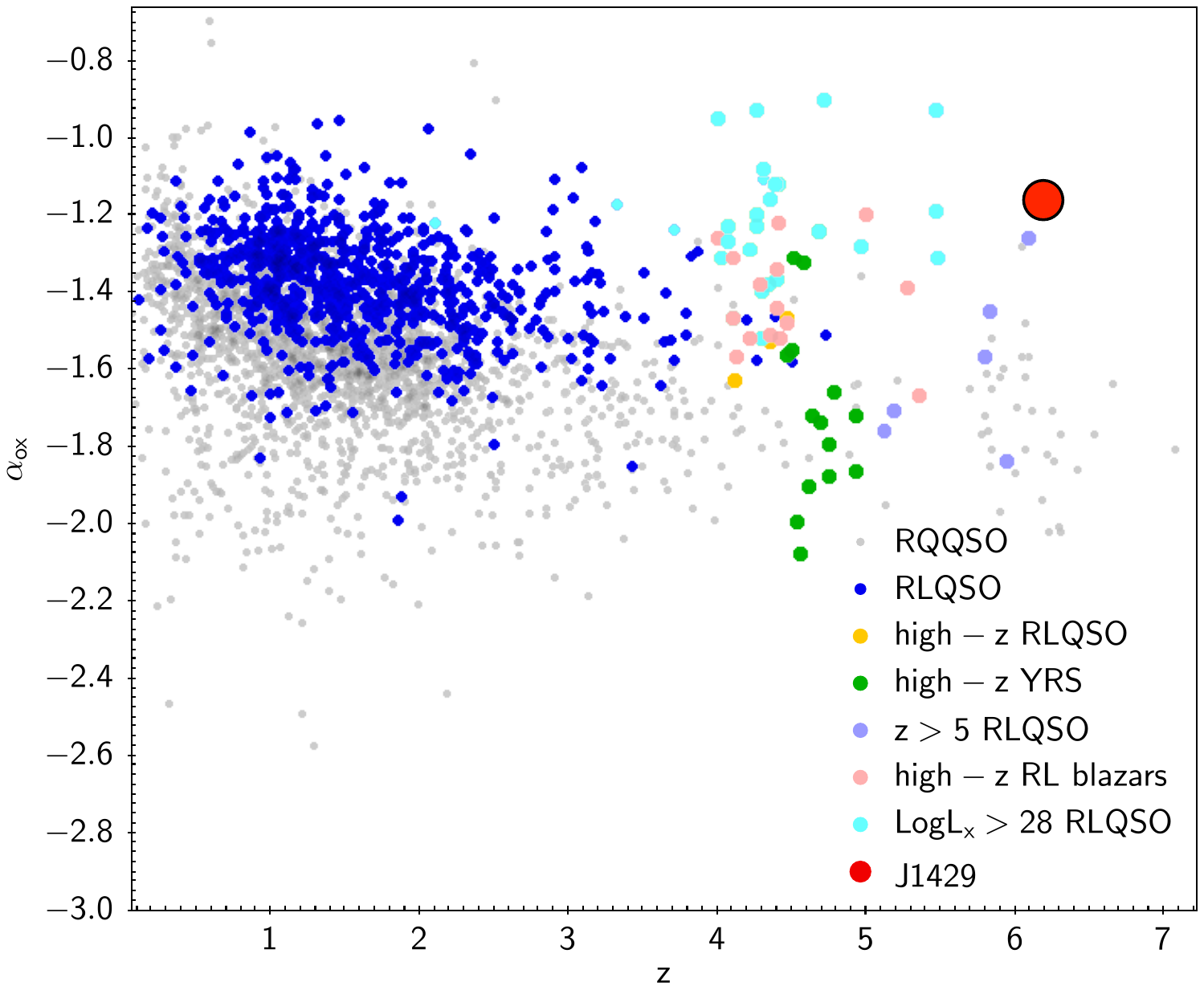}
	\includegraphics[trim=2.cm 0.cm 2.2cm 0cm, clip=true,width=0.95\columnwidth]{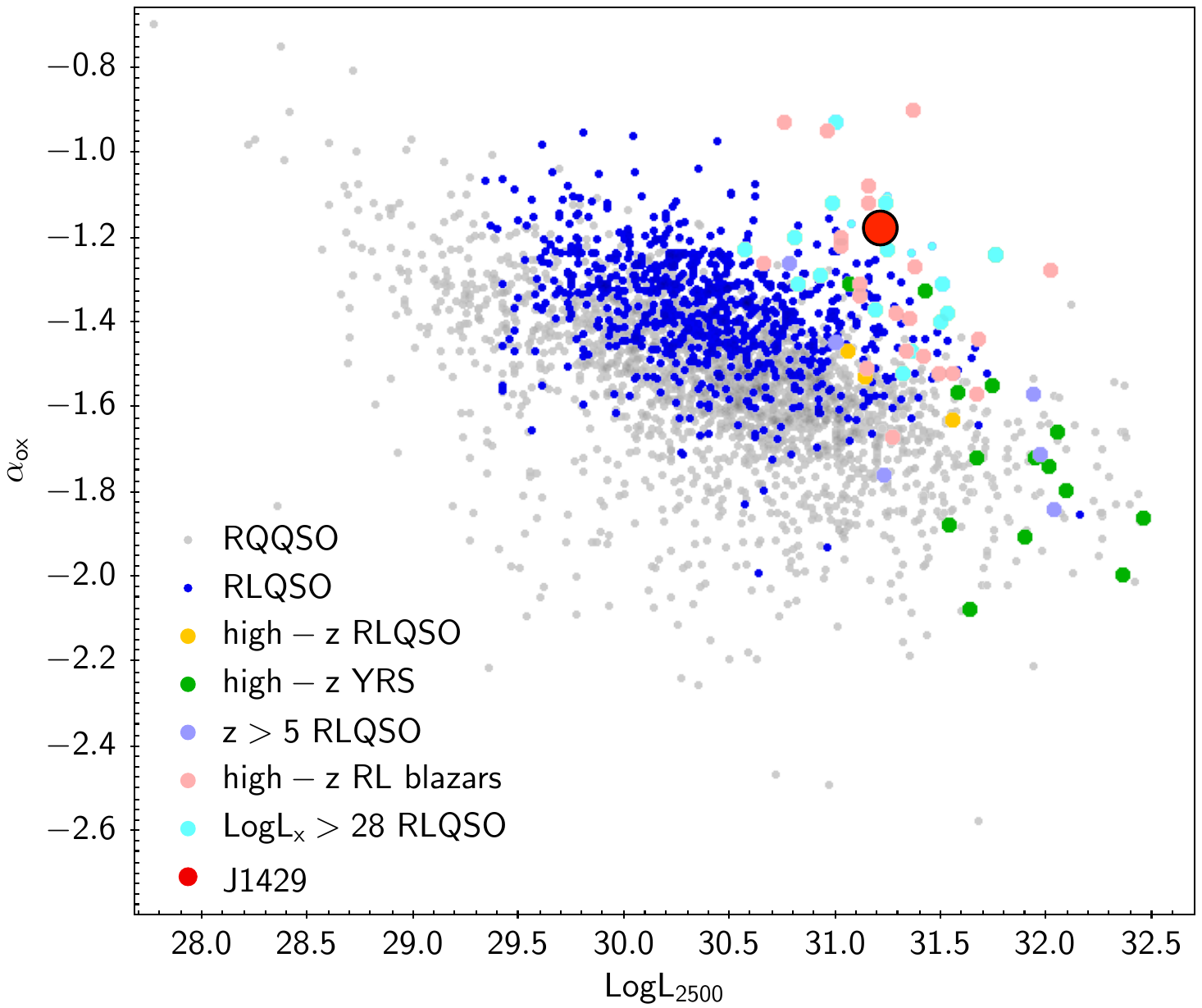}
	\includegraphics[trim=0.cm 2.cm 0.cm 2.2cm, clip=true,angle=90,width=0.95\columnwidth]{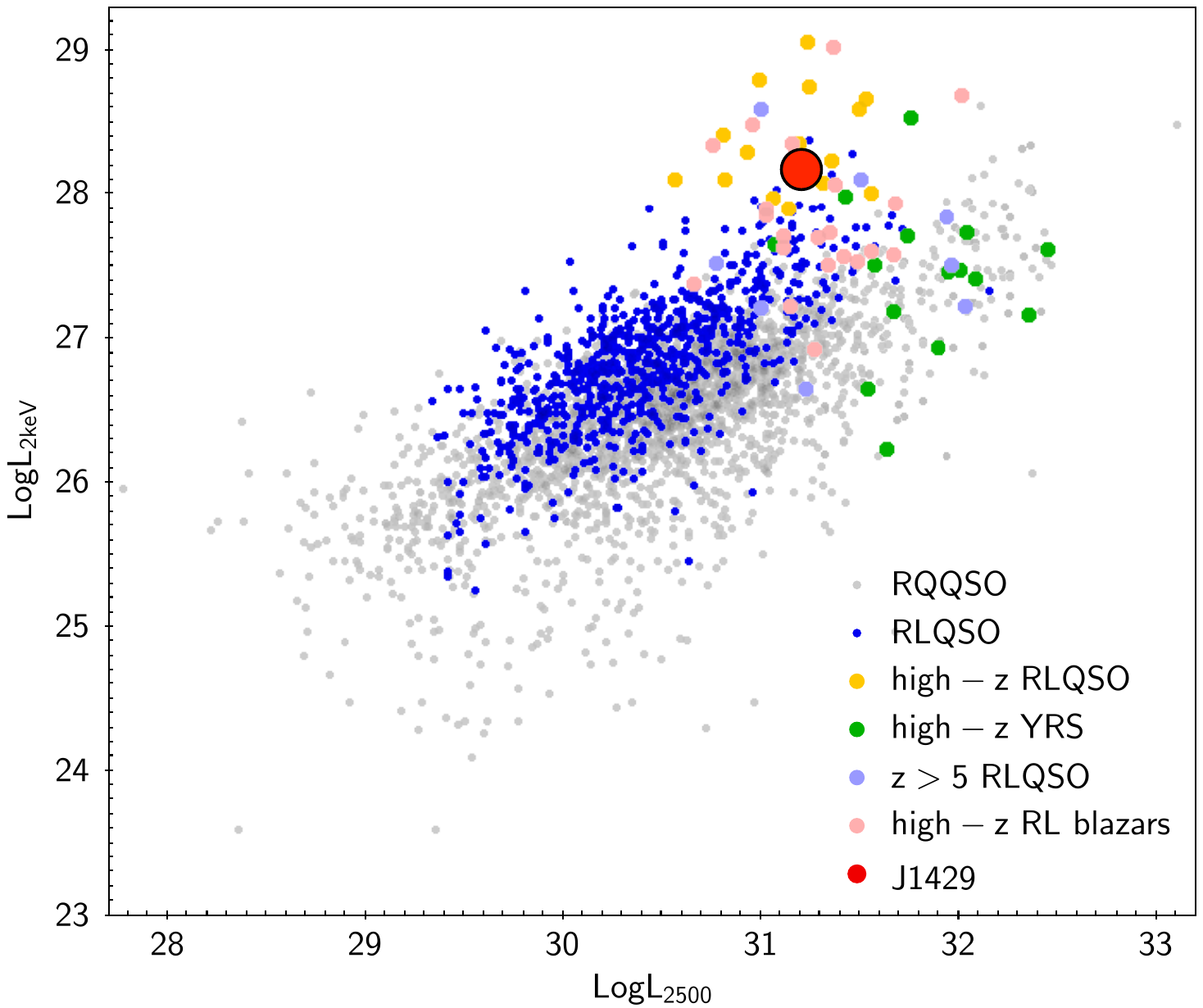}
	\includegraphics[trim=0.cm 2.cm 0.cm 2.2cm, clip=true,angle=90,width=0.95\columnwidth]{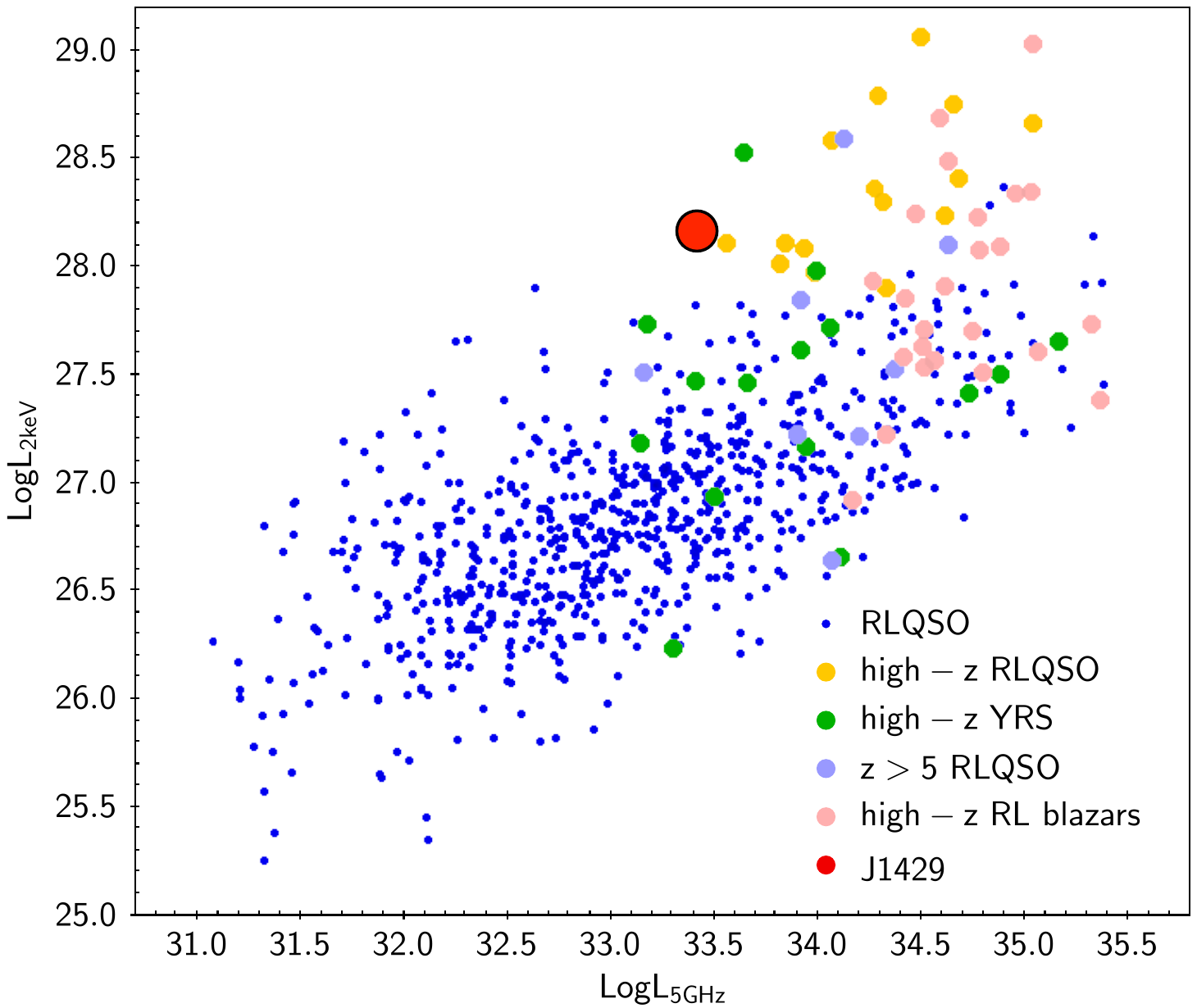}
	\includegraphics[trim=0.cm 2.cm 0.cm 2.2cm, clip=true,angle=90,width=0.95\columnwidth]{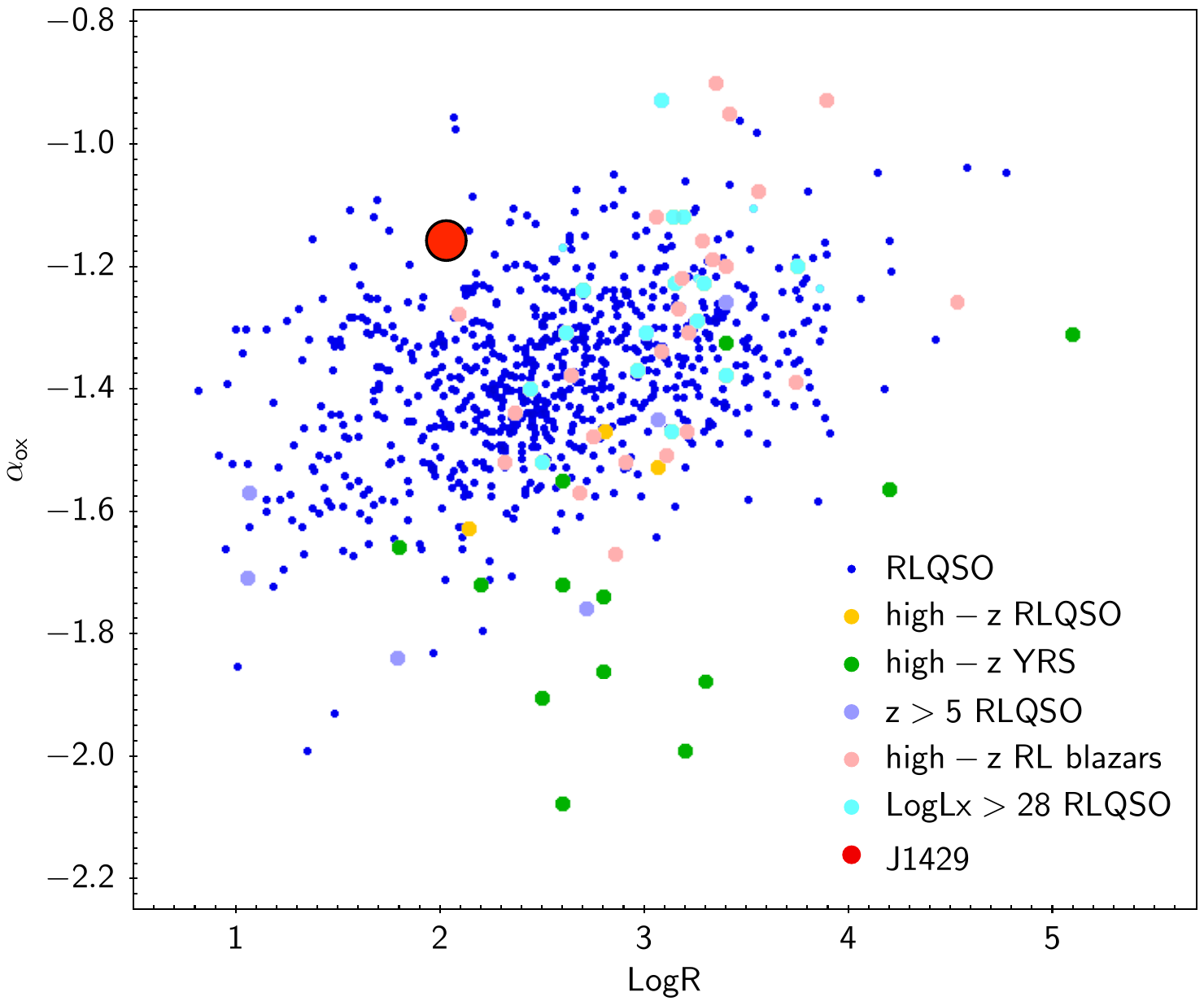}
	\includegraphics[trim=0.cm 2.cm 0.cm 2cm, clip=true,angle=90,width=0.95\columnwidth]{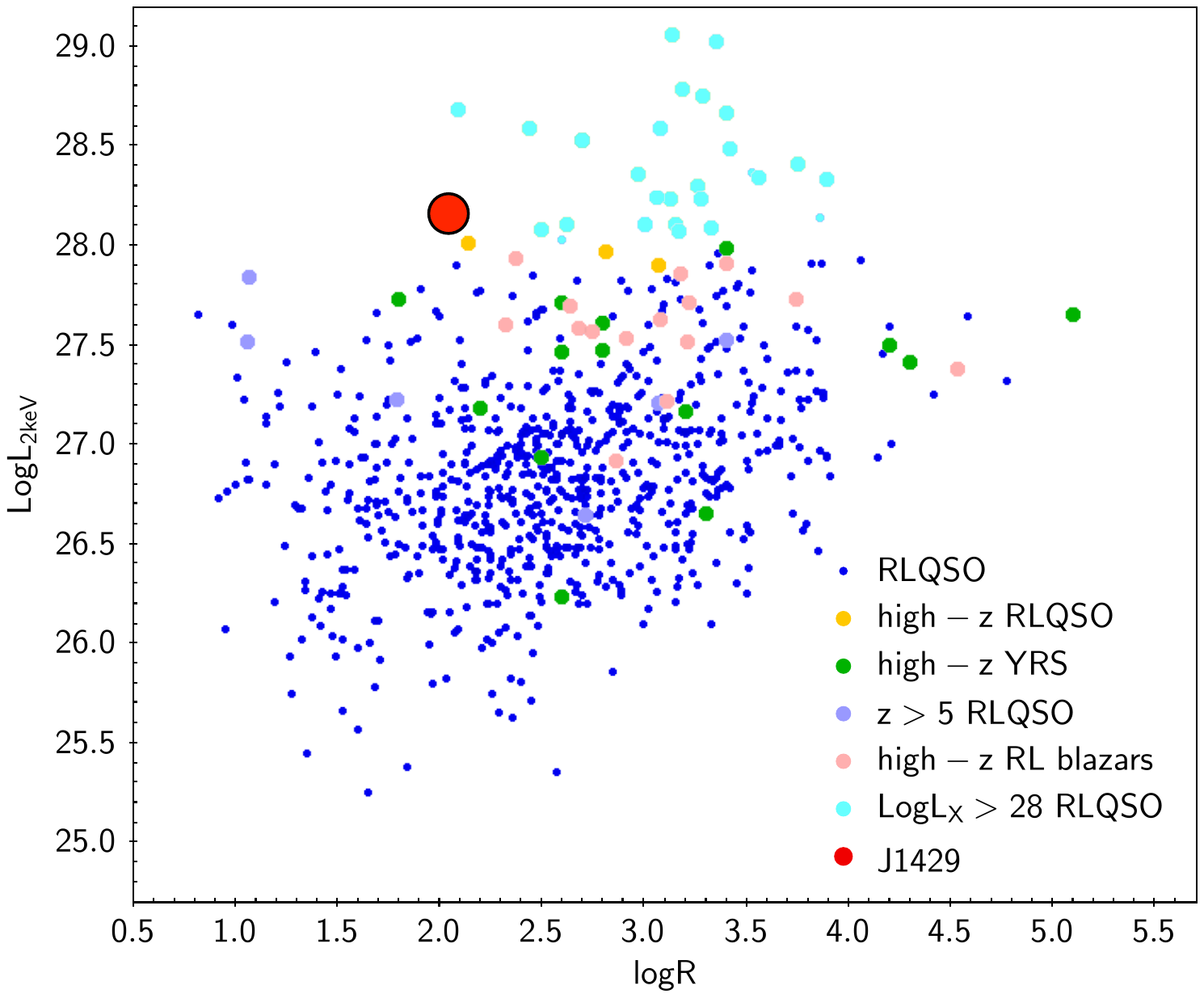}
    \caption{Comparison plots for \gal. Top: optical--to--X-ray power-law slope $\alphaox{}$ versus redshift (left) and versus the 2500 \AA\, luminosity density (right). Middle: 2\,keV luminosity density versus 2500 \AA\, luminosity (left) and versus 5\,GHz luminosity density (right). Bottom: $\alphaox{}$ versus radio loudness (left) and 2\,keV luminosity density versus radio loudness (right). All quantities are computed in the quasars' rest frames. The RQ quasars (grey dots) are from \citet{Shemmer2006}, \citet{Just2007}, \citet{Lusso2016}, \citet{Martocchia2017}, \citet{Nanni2017} and \citet{Vito2019b}. The samples of radio-loud quasars are taken from \citet{zhu2019} (yellow dots) and \citet{zhu2020} (blue dots), the high-redshift blazars (pink dots) from \citet{Ighina2019}, the X-ray sample of young radio sources at $z>4.5$ (green dots) from \citet{Snios2020}; $z>5$ radio-loud quasars with reported X-ray detections are marked with violet dots  \citep[see][]{Khorunzhev2021}. In all the panels but the middle ones, for each sample we highlighted in cyan the radio-loud quasars  with the 2\,keV luminosity densities comparable with, or higher than, that of \gal\, ($L_{\rm 2\,keV}\gtrsim 10^{28}$\,\lumcgs\,Hz$^{-1}$). In the top panels, we have also included for a comparison the samples of radio-quiet quasars from \citet{Shemmer2006}, \citet{Just2007}, \citet{Lusso2016}, \citet{Martocchia2017}, \citet{Nanni2017} and \citet{Vito2019b}. }
    \label{fig:diagnostic}
\end{figure*}

\section{Discussion}
\label{sec:disc}

The \chandra{} observation confirms the high X-ray luminosity of \gal\ (the  extrapolated rest-frame 0.1-100\,keV luminosity $\sim 4\times 10^{46}$\,\lumcgs) and returns a photon index value within the typical range of high-$z$ quasars  \citep[$\Gamma\sim 1.5-2.2$, e.g.][]{Vito2019b,zhu2019}. 
In order to investigate the origin of the observed X-ray emission, in Figure~\ref{fig:diagnostic} we compared the radio, optical and X-ray properties of our target (in Table \ref{tab:plotpar}) with those of samples of quasars from the literature. For this purpose, we used the estimated radio-loudness parameters and the X-ray--to--optical (2500 \AA) luminosity ratios, expressed as spectral slopes between 2500 \AA\, and 2 or 10\,keV, namely $\alpha_{\rm ox}=0.3838 \log[L_{\rm 2\,keV}/L_{2500}]$ \citep{Tananbaum1979} and $\tilde{\alpha}_{\rm ox}=0.3026 \log[L_{\rm 10\,keV}/L_{2500}]$ \citep{Ighina2019}, where $L_{2500}$, $L_{\rm 2\,keV}$ and $L_{\rm 10\,keV}$ are the corresponding rest-frame luminosity densities in units of \lumcgs\ Hz$^{-1}$. The rest-frame luminosities of \gal{} at 2 and 10\,keV, $7.7\times 10^{45}$\,\lumcgs, give $\alphaox{}=-1.15$ and $\tilde{\alpha}_{ox}=-1.12$, which are both in a good agreement with the previous estimates by \citet{medvedev20}, while the radio loudness is $R=109\pm9$ \citep{Banados2015}.

{\bf
\begin{table*}
	\caption{\gal{} rest-frame luminosity densities.
	\label{tab:plotpar}}
	\begin{tightcenter}
	\footnotesize
	\begin{tabular}{c c c c c }
		\hline \hline
		 $L_{\rm 5\,GHz}$ & $L_{4400\angstrom}$  & $L_{2500\angstrom}$  &$L_{\rm 2\,keV}$   &$L_{\rm 10\,keV}$ \\
		 \lumcgs{}Hz$^{-1}$ &\lumcgs{}Hz$^{-1}$ &\lumcgs{}Hz$^{-1}$ &\lumcgs{}Hz$^{-1}$ &\lumcgs{}Hz$^{-1}$ \\
\hline		 
		      $(3.1\pm0.1)\times 10^{33}$  &$(2.8\pm0.2)\times 10^{31}$            &$(1.6\pm0.2)\times 10^{31}$              &$(1.6^{+0.8}_{-0.6})\times 10^{28}$   &$(3.2^{+0.9}_{-0.7})\times 10^{27}$  \\
		\hline
	\end{tabular}
    \end{tightcenter}
    {Notes: the values of $L_{\rm 5\,GHz}$ and $L_{4400\angstrom}$ are taken from \citet{Banados2015}. $L_{4400\angstrom}$ was calculated from the WISE W1 magnitude. $L_{2500\angstrom}$ is taken from \citet{medvedev20} and was estimated using the median
composite SED of radio-loud quasars from \citet{Shang2011} normalized to the observed y-band flux density from the Pan-STARRS1 survey \citep{Chambers2016}. The X-ray flux density are from the analysis here presented.}
\end{table*}
}
For the comparison, we used the samples of radio-loud quasars from \citet{zhu2020} and $z>4$ radio-loud quasars from \citet{zhu2019}, high-redshift blazars presented in \citet{Ighina2019}, and the sample of young radio sources at $z>4.5$ from \citet{Snios2020}. We also collected from the literature information on $z>5$ radio-loud quasars with reported X-ray detections, not present in the previously mentioned samples \citep[see][and references therein]{Khorunzhev2021}. 
In addition, in the $\alphaox${} vs. $z$ panel, as well as the $\alphaox${} vs. $L_{2500}$ panel of Figure~\ref{fig:diagnostic}, we included the samples of radio-quiet quasars from \citet{Shemmer2006}, \citet{Just2007}, \citet{Lusso2016}, \citet{Martocchia2017}, \citet{Nanni2017} and \citet{Vito2019b}, which, although collectively not complete, ensure a good redshift coverage.

Looking at the plots in Figure~\ref{fig:diagnostic}, we can make some basic considerations. The $\alphaox{}$ of \gal\ is relatively high, in particular in comparison with the high-$z$ RQ and RL quasars. Although we confirm that \gal\ does not follow the anti-correlation between $\alphaox$ and $L_{2500}$ known for lower-redshift AGN \citep{Lusso2016}, the deviation from the relation is less extreme than what reported in \citet{Wolf2021} based on the \xmm~data (see their Figure~6). The $\alphaox-L_{2500}$ and $L_{\rm 2\,keV}-L_{2500}$ panels confirm that the reason for this is an excess in X-rays rather than a deficit in the UV luminosity. As noted by \citet{medvedev20}, the X-ray luminosity of \gal\ is comparable with the most radio-loud ($\log R\gtrsim 2.5$) high-$z$ sources in the sample of \citet{zhu2019}, which are supposed to have a significant contribution in X-rays by a Doppler-boosted jet emission. However, this explanation is to some extent contradicted by a relatively low $R$ value for our target, which is in the lowest tail of the samples of \citet{zhu2019} and \citet{Ighina2019}, though $R$ may not be an ideal tracer of a jet-activity in high-$z$ quasars accreting at the highest rates \citep[see][]{Sbarrato2021}. Moreover, the average photon index of the $z>4$ blazar sources is markedly flatter \citep[$\Gamma=1.4$; ][]{Ighina2019} than our revised value. In fact, referring to the $\tilde{\alpha}_{ox}$ vs. $\Gamma$ classification plot proposed by \citet{Ighina2019}, the steep photon index locates \gal\ among the non-blazar sources. Note also that, while the brightest X-ray (say $L_{\rm 2\,keV}> 10^{28}$ \lumcgs{}Hz$^{-1}$) RL quasars are radio bright (at 5 GHz), the opposite is not true, i.e. high $L_{\rm 5\,GHz}$ values do not necessarily imply high $L_{\rm 2\,keV}$ (see the middle right panel in Figure~\ref{fig:diagnostic}), as $L_{\rm 5\,GHz}$ could be the sum of beamed (jet) and unbeamed (lobes) radio emission.  

\citet{zhu2020} investigated the origin of the X-ray excess of RL vs. RQ quasars using a large sample of optically-selected RL quasars and concluded that, only in the case of flat-spectrum radio quasars (FSRQs), this excess is due to a direct contribution of a boosted jet emission. For the majority of steep-spectrum radio quasars (SSRQs) instead, the authors argue that the radio, optical and X-ray parameters point to a disk corona origin of the X-ray emission, although there must be a physical link between the coronal and the jet activity, which is manifested through the increase of the X-ray emission as a function of the radio-loudness. On one hand, our target seems to fit well into this picture, in view of its photon index value in line with the corona emission. On the other hand, its X-ray luminosity exceeds that of SSRQs.  One could argue that the production of the X-ray emission in the disk corona of \gal\ is, for some reasons, more efficient than in the cases of SSRQs from the Zhu et al. sample. A caveat for this comparison is indeed that the sample of \citet{zhu2020} does not include SSRQs at the same redshift as \gal. 

\citet{Vito2019b} investigated the X-ray properties of a sample of $z>6$ RQ quasars and did not find evidence for an evolution of the disk/hot corona structures with respect to the lower-redshift counterparts. If so, the corona-related excess X-ray emission would be unique to SSRQs. To test this possibility, one would need to significantly increase the number of $z>6$ SSRQs. Incidentally, we note that \cite{Shen2019} classifies \gal{} as a weak emission-line QSO \citep[WLQSO, where the definition is based on the rest-frame equivalent width of C IV $<$15.4 \AA, ][]{Fan1999,Diamond-Stanic2009}. The proposed explanations for this class of objects involve young accreting systems or different accretion and absorption conditions in the innermost region of the QSO \citep[e.g.][]{Shemmer2010,Laor2011,Luo2015}. 

Alternatively, it is possible that multiple radiative components contribute to the total X-ray emission of \gal. \citet{medvedev21} explored the IC/CMB scenario for the X-ray emission. While part of their reasoning was based on the steep X-ray photon index measured by \xmm, which we now revised, a contribution of the jet via IC/CMB to the total X-ray emission remains a possibility. We exploited the available \chandra\ data to search for an extended emission on angular scales $\gtrsim 0.5\arcsec$, corresponding to a physical scale of $\sim 2.8$\,kpc. 
The resolved kpc-scale X-ray quasar jets are typically characterized by the jet-to-core luminosity ratios $R_{\rm jc}\sim 2\%$ \citep{Marshall2018}, and even for the most luminous jets at high redshift  $R_{\rm jc}\lesssim 10\%$ \citep{Siemiginowska2003,Cheung2006,Schwartz2020,Ighina2022c}. For our 30\,ksec \chandra\ observation, this would give a maximum of 10 net counts in the most optimistic scenario ($R_{\rm jc}\sim 10\%$, assuming a standard photon index $\Gamma=1.7$), while for $R_{\rm jc}\sim 2\%$ the jet X-ray emission would be below the detection limit.

Based on the available X-ray dataset, we place an upper limit $9\times 10^{44}$\,\lumcgs to any X-ray component on scales $>1.5\arcsec$ ($> 8$\,kpc), thus excluding the presence of a luminous, jet-related emission far outside of the galactic host. The putative count excess that we measure in the 0.5\arcsec--1.5\arcsec annulus ($\sim 3-8$\,kpc projected scale) implies an X-ray luminosity $\sim 4\times 10^{45}$\,\lumcgs. If due to a kiloparsec jet, this would make it for a remarkable $\sim 20\%$ of the total observed X-ray flux, well beyond the $R_{\rm jc}$ observed ranges. 
Even in this case, however, the $>3$\,kpc jet would be far from being the dominant X-ray contribution. 

The imaging analysis leads us to conclude that the bulk of the X-ray emission must be produced within the central $\sim 3$\,kpc (projected) region. Nonetheless, Doppler boosting would still be needed to explain the high X-ray luminosity in terms of the IC/CMB jet emission. \cite{Shen2019} argue that the weakness of the C IV line could be due to contamination of the UV continuum by the jet emission, a possibility taken into consideration also for the far IR (FIR) emission \citep{Khusanova2022}. Therefore, a blazar-like nature of the source cannot be fully ruled out, although it stands at odds with the established radio properties. In this scenario, the high-energy emission could be produced via inverse-Compton scattering of the nuclear photons (UV and IR photons) in the inner segment of a relativistic jet. Follow-up multi-wavelength observations could probe this hypothesis by searching for a variability of the emission, or lack of thereof.

We also briefly consider in this context the young radio source scenario. High-energy emission is predicted to be produced in the compact lobes of young radio galaxies \citep{Stawarz2008b} and in the jets of young radio quasars \citep{Migliori2014}. Support to a non-thermal, high-energy component in young radio sources has come from the detection of a handful of these sources in the $\gamma$-ray band with the {\it Fermi} telescope \citep{Migliori2016,Fermi2020,Principe2020,Principe2021}. However, modeling of the broad-band high-energy output of the compact lobes of young radio galaxies points to much lower X-ray luminosities than \gal, namely $\sim 10^{41}-10^{42}$\,\lumcgs \citep[see, e.g., the SED modeling of the $\gamma$-ray detected young radio galaxy PKS\,1718$-$649,][and references therein]{Sobolewska2022}. The high-energy emission of young radio quasars can be as high as $\sim 10^{45}-10^{46}$\,\lumcgs{} but it is typically charaterized as variable, suggesting a blazar-type origin \citep{Siemiginowska2008,Principe2021}.

To conclude, the bulk of the X-ray luminosity of \gal\ seems either originating in the quasar accretion or in the (aligned?) jet,  with the latter hypothesis being disfavored by the source's radio properties. As discussed in \cite{Khusanova2022}, X-ray dominated regions (XDR) produced by the X-ray photons from the accreting AGN can contribute to the observed [CII] emission. If we use the relation between [CII] line and the 2--10\,keV luminosity for XDR $L_{\rm [CII],\,XDR}=2\times 10^{-3} L_{\rm 2-10\,keV}$ \citep{Stacey2010}, we obtain $L_{\rm [CII],\,XDR}\sim 2\times 10^{43}$\,\lumcgs, a value comparable with the observed $L_{\rm [CII]}$ \citep{Khusanova2022}. Indeed, this is a rough estimate, as it is unlikely that the whole $L_{\rm [CII]}$ is produced by XDR \citep[see][for a review]{WVC2022}. However, it shows that, in this system, XDR could be in principle an important contribution to gas heating in addition to the starlight.

\section{Conclusions}
\label{sec:conc}

We presented the results of the $\sim 30$\,ksec \chandra\ observation of the high-$z$ radio quasar \gal. The high angular resolution of \chandra\ allowed us to identify the X-ray sources in the field of \gal, and to derive the X-ray spectrum of the target free of the contaminating source, which was not spatially resolved in the previous eROSITA and \xmm\ observations. In addition, we were able to place constraints on the X-ray emission of a putative kiloparsec jet, concluding that the bulk of the X-ray emission must be produced within a $\sim 3$\,kpc-radius central region, either in the disk-corona system, or in the jet. In the former case, the accretion luminosity of \gal\ appears higher than that of similar systems observed at high redshifts, such as steep spectrum radio quasars, and could significantly impact the ISM. In the latter case, the non-thermal emission should be boosted, implying a (moderately) aligned jet as in blazar sources. 

The analysis of the \chandra\ image pointed to a count excess over the PSF predictions in the 0.5\arcsec--1.5\arcsec\, central region, corresponding to a high X-ray luminosity ($\sim 4\times10^{45}$\,\lumcgs). While a deeper \chandra\ observation is needed to confirm this result, we mention the possibility that --- instead of being related to \gal{} (i.e., a kpc-scale jet) --- this excess  could be revealing of a separate X-ray source. This is an intriguing hypothesis given the observational evidences that \gal\ resides in a merging system \citep{Khusanova2022}.

The case of \gal\ effectively shows how wide-field, large-effective area X-ray telescopes are key for the discovery of high-$z$ quasars, while high-angular resolution observations are needed to ensure the correct characterization of the X-ray emission.

\section*{Acknowledgements}

The authors would like to thank the referee for useful suggestions and comments. Support for this work was provided by the National Aeronautics and Space Administration through \chandra\ Award Numbers GO8-19093X and GO0-21101X
issued by the Chandra X-ray Observatory Center, which is operated by the Smithsonian Astrophysical Observatory for and on behalf of the National Aeronautics Space Administration under contract NAS8-03060. G.M. acknowledges financial support from INAF mini-grant "The high-energy view of jets and transient" (Bando Ricerca Fondamentale INAF 2022).
A.S., M.S., D.A.S. and V.L.K. were supported by NASA contract NAS8-03060 (Chandra X-ray Center).
C.C.C. was supported at NRL by NASA DPR S-15633-Y. {\L}.S. was supported by the Polish NSC grant 2016/22/E/ST9/00061. This research has made use of data obtained from the Chandra Data Archive, and software 
provided by the Chandra X-ray Center (CXC) in the application packages CIAO and Sherpa.

\section*{Data Availability}

All data underlying this article are already publicly available from NASA's HEASARC archive (\url{https://heasarc.gsfc.nasa.gov/}), \chandra’s  Data Archive (\url{https://cxc.harvard.edu/cda/}).



\bibliographystyle{mnras}
\bibliography{j1429,all_references} 



\bsp	
\label{lastpage}
\end{document}